# Interlayer charge transfer in graphene–2D polyimide heterostructures

Francesca Falorsi[1*], Shuangjie Zhao[2*], Kejun Liu[2,3], Christian Eckel[1], Jonas F. Pöhls[1], Wiebke Bennecke[1], Marcel Reutzel[1], Stefan Mathias[1,4], Kenji Watanabe[5], Takashi Taniguchi[6], Zhiyong Wang[2,7], Miroslav Polozij[2,8,9], Xinliang Feng[2,7], Thomas Heine[2,8,9,10], R. Thomas Weitz[1,4]

[1] I. Institute of Physics, Georg-August-University of Göttingen
[2] Faculty of chemistry and food chemistry, Technical University of Dresden
[3] Functional Nano & Soft Materials (FUNSOM), Jiangsu Key Laboratory for Carbon-Based Functional Materials & Devices, Soochow University, Suzhou 215123, China
[4] International Center for Advanced Study of Energy, Göttingen ICASEC5
[5] Research Center for Electronic and Optical Materials, National Institute for Materials Science, 1-1 Namiki, Tsukuba 305-0044, Japan
[6] Research Center for Materials Nanoarchitectonics, National Institute for Materials Science, 1-1 Namiki, Tsukuba 305-0044, Japan
[7] Max Planck Institute of Microstructure Physics, 06120 Halle (Saale), Germany
[8] Helmholtz-Zentrum Dresden-Rossendorf, HZDR, Bautzner Landstr. 400, 01328 Dresden, Germany.
[9] Center for Advanced Systems Understanding, CASUS, Untermarkt 20, 02826 Görlitz, Germany.
[10] Department of Chemistry, Yonsei University, Seodaemun-gu, Seoul, 120-749 Republic of Korea.
* F. Falorsi and S. Zhao contributed equally.

The vertical integration of multiple two-dimensional (2D) materials in heterostructures, held together by van der Waals forces, has opened unprecedented possibilities for modifying the (opto-)electronic properties of nanodevices. Graphene, with its remarkable opto-electronic properties, is an ideal candidate for such applications. Further candidates are 2D polymers, crystalline polymeric materials with customizable structure and electronic properties that can be synthesized in all mathematically possible Bravais lattices. In this study, we investigated the optoelectronic properties of a heterostructure created by pristine graphene and a rectangular 2D polyimide (2DPI) film. This imprints a new superlattice on graphene in conjunction with a direct influence on its electronic properties. Theoretical and experimental analyses reveal that interlayer charge exchange between the 2D polymer and graphene induces hole doping in the graphene layer. We have also observed that the properties of the heterostructure are dependent on the substrate used in experiments, likely due to the porous character of the 2DPI allowing direct interaction of graphene with the support. These findings highlight the unique ability to tailor functionalities in 2D polymers-based heterostructures, allowing the development of optoelectronic devices with precisely engineered properties and stimulating further exploration of the diverse phenomena accessible through tailored designs of the 2D polymers.

In the past decades, the rise of graphene [1, 2] and other two-dimensional (2D) materials has offered an unprecedented possibility to combine them to layer-stacked heterostructures (HS) held together by van der Waals (vdW) forces. Strong interlayer effects enable access to new and interesting physical phenomena [3]. Prominent examples include the band gap opening in bilayer graphene (BLG) [4], the indirect-to-direct band gap transition in transition metal dichalcogenides [5–7], and the semiconductor-to-metal transition in noble metal dichalcogenides [8, 9]. However, graphene and other 2D crystals have fixed lattice structures that are limited in their tunability and modification. To increase structural flexibility and enable functional designs, 2D polymers and their layer-stacked variant, 2D covalent organic frameworks (COFs), have been developed for HS in recent years [10–15]. 2D polymers are crystalline, layered materials with organic building blocks connected laterally via covalent bonds with monolayer or few-layer (<10) thickness. Notably, 2D polymers can form all mathematically possible 2D lattices (e.g., hexagonal, square, kagome)[16, 17], directly influencing their electronic band structure, which can be metallic, semiconducting, or exhibit topological bands, Dirac points, and flat bands. Additionally, their chemical composition is highly modifiable, affecting the work function and interlayer charge exchange, thereby altering the density of states. Having these novel materials with tunable intriguing properties, it is expected that when coupled with graphene, novel physical phenomena can be induced into graphene.

In this work, we investigate the electrical and spectroscopic properties of a heterostructure (HS) of monolayer graphene with a 2D rectangular polyimide (Figure 1a) (denoted below as 2DPI/graphene). The ultrafast interlayer charge transfer between this 2DPI and chemically exfoliated graphene has already been shown by some of us [13]. Here, we further investigate these HSs by combining high-quality exfoliated graphene with 2D polyimide film of different thicknesses. This approach allows us to analyze both the interlayer charge transfer in the HS as well as the impact of the 2DPI on the charge transport within the graphene layer. As we detail below, the 2D polymer induces hole doping in the graphene, and the strength of the doping increases with the layer number of the polymer. In our analysis we have utilized a large number of complementary methods, not only to give a full characterization of the HSs but also to give reference data for future HSs of the similar type.

We first extensively utilize theoretical models to explain the electronic properties of the 2DPI/graphene HS. Standard theoretical models of layered materials, which only consider standalone layers in vacuum, are not suitable for the 2D polymer/graphene system. This stems from the porous nature of the 2DPI with pores large enough for the graphene to interact through the 2DPI layer with the substrate under it. To take this effect into account, we have investigated three different models; i) standalone 2DPI/ graphene, ii) 2DPI/graphene deposited on $SiO_2$ as in transport experiments, using 2D $SiO_2$ [18] as a substrate model, denoted $SiO_2$/2DPI/graphene, iii) 2DPI/graphene deposited on Si as in ARPES experiments, using a H-terminated 4-layer 2D model of Si(100) surface as a substrate model, denoted Si/2DPI/graphene (Figure 1b).

There are two important structural effects influencing the graphene in the HS with the porous 2D polymer. Due to the van der Waals interaction between the 2DPI and graphene, a superlattice forms in graphene mirroring the lattice structure of the polyimide. This is manifested by the corrugation of graphene towards the 2DPI pore. In the case of standalone 2DPI/graphene HS, the corrugation amplitude is about 1.12 Å, however, with Si and $SiO_2$ substrates, the maximum deformation of graphene towards the support is 1.5 Å and the total corrugation amplitude is almost 3 Å (Figure 1c). The second structural effect on graphene comes from the relative rotation of the graphene and 2D polyimide layers. Experimentally, such an angle is very hard to control due to the polycrystalline nature of the 2DPI used and should be considered random. This causes the superlattice imposed on graphene to be in different orientations with respect to the graphene lattice. To investigate this, we have tested several rotation angles (0°, 4°, 12.6° and 21.6°), which produce models with small enough unit cells to study their electronic properties.

In the case of the standalone 2DPI/graphene, the HS band structure is a simple superposition of the 2D polyimide and graphene band structures, irrespective of the rotation angle (Figures S1, S2). More pronounced electronic changes can be seen with the introduction of a substrate ($SiO_2$ or Si). Due to the size of the systems, we were not able to include the substrate directly in most electronic properties calculations. Instead, we only used it to obtain the HS geometry and removed the substrate for the band structure calculation. This does not bring any significant error; see the example of Si/2DPI/graphene in Figure S3. The introduction of a $SiO_2$ substrate does not bring any changes to the 2DPI/graphene band structure (Figure S4). On the other hand, with the introduction of a Si substrate, a gap of 7.5 meV opens at the Dirac cone, independent of the HS rotational angle (Figures 1d, S6). A more significant change is seen in the density of states (Figures 1e, S4, S5) for both the $SiO_2$ and Si supported HS, where two distinct peaks emerge below and above the Fermi level, instead of a single peak above the Fermi level in the standalone HS. This relates to the difference in charge transfer between the 2D polyimide and graphene (Figures 1f, S11), which is much stronger in the corrugated structures of the substrate-deposited HS.

**Electronic properties of 2DPI/graphene heterostructures**

To experimentally analyze the optoelectronic properties of the 2DPI/graphene HS, we first focus on the $SiO_2$ supported system. This brings the additional advantage that the $SiO_2$ is insulating and can be used as dielectric in a field-effect transistor geometry allowing to electrostatically tune the Fermi level in the HS. The charge transfer between 2D polyimide films of different thicknesses and graphene was verified with various complementary experimental approaches: two scanning probe methods (scattering-type scanning near-field optical microscopy (SNOM) and Kelvin Probe Force Microscopy (KPFM)) and electrical measurements. Multiple samples were studied, where exfoliated graphene flakes were deposited onto 2D polyimide films via a dry transfer method (described in [19]). The monolayer 2DPI was synthesized as described by our previous report [12] using the Langmuir-Blodgett method and has a homogeneous thickness of 0.8 nm [13]. To facilitate the comparison between the HS configuration and bare graphene, a portion of the graphene had been stamped onto $SiO_2$ so that the optical properties of bare graphene and the HS can be compared.

A typical image of a $SiO_2$/2DPI/graphene HS is shown in Figure 2a. Here, the HS is mapped using SNOM, where in addition to the topographic information we can record local differences in optical conductivity. The relative optical contrast between the HS and the bare graphene is consistent with a hole doping of graphene by the 2D polyimide, as we describe in more detail below. The near-field optical signal is directly linked to the amplitude and phase of the electromagnetic field inside the nanogap between the tip and the sample and is thus related to the complex optical conductivity of the sample [20]. Figure 2a shows the 3$^{rd}$ harmonic optical amplitude taken with a laser excitation energy of around 115 meV, where it is apparent that the optical amplitude in graphene is larger compared to the HS region. To analyze the doping of graphene, we also have studied the excitation-energy dependent relative contrast between 115 meV and 134 meV (see Figure S5), where the contrast between the HS and the bare graphene diminishes with increasing excitation energy. Figure 2b summarizes this trend where the difference in contrast of the gold normalized 3rd harmonic optical amplitude of the HS and the graphene (sample depicted in Figure S6a) is shown as a function of excitation energy. This observation is consistent with the density-dependent optical conductivity of graphene, where we expect a transition from Drude optical response at lower excitation energies to inter-band transitions at higher excitation energies (see inset Figure 2b) [21, 22]. At constant doping the transition between the two different absorption mechanisms will occur above a particular excitation wavelength. This is the reason that, for lower excitation energies, there is a large contrast between the HS (Drude response dominates) and graphene (inter-band processes contribute), while at high excitation energies, we find a doping-independent optical response (both in HS and graphene-inter-band processes dominate). Therefore, the lower contrast of the HS at low excitation energies is an indication for its hole doping compared to the bare graphene [23–26]. Finally, we note that the SNOM images of the HS appear very homogeneous, which implies that at the scale of the resolution (around 50 nm) the doping is uniform.

A complementary picture of the Fermi energy difference between the graphene and HS can be obtained with KPFM [27]. In Figure 2c, the KPFM image of the region highlighted by the dotted rectangle in Figure 2a is shown. The surface potential measured ($U_{SP}$) by the KPFM is related to the work function difference between the measuring tip and the sample [28, 29]. The mean $U_{SP}$ values are obtained by fitting a Gaussian curve to the $U_{SP}$ histogram of the graphene and HS regions (Figure 2d with an error equal to the standard deviation (σ) of the fit). By performing the weighted average over 6 different images acquired with different scanning directions on the same sample, we obtained an averaged $U_{SP}$ difference between the HS and the graphene of $\Delta V^{mono}_{HS-Gr}$= (41.6±4.1) mV. Since in graphene the $U_{SP}$ is directly connected to the doping level of the system [30, 31] via the relation $\boldsymbol{\varepsilon_F = \hbar\, v_F \sqrt{\pi\, n}}$ ( where $\boldsymbol{v_F}$ represents the Fermi velocity of the graphene monolayer equal to $\boldsymbol{v_F}$ =10$^6$ m/s) [32] it is estimated that the 2DPI induces a hole doping density of $\Delta n_{mono}$=(1.27± 0.15) 10$^{11}$ cm$^{-2}$ in the graphene layer of the HS shown in Figure 2a.

In general, the interaction between 2DPI and graphene can go beyond pure hole doping, and also the charge carrier mobility can be distinct in the HS compared to bare graphene, or the 2DPI might open a bandgap in graphene (due to the breaking of sublattice symmetry in graphene). This is why we have performed direct electrical transport measurements in various samples. In Figure 2e a gate sweep of the HS shown in Figure S6a is presented and compared to a gate sweep of the same graphene flake which is not in contact with the 2DPI. From the relative position of the charge neutrality point of the two devices, we are able to identify the relative doping level. Consistent with the scanning probe measurements, the 2DPI induces hole doping in the graphene; in this case of Δn = (1.55±0.01) x10$^{12}$ cm$^{-2}$. The calculations are performed as described in the electrical measurement paragraph of the Methods section. The relative doping densities are different in the transport and KPFM measurements, likely due to sample-to-sample variations. The 2DPI also induces additional scattering [35] as we can identify by a decrease of the charge carrier mobility from 1580 to 901 cm²/Vs (at $10^{12}$ cm$^{-2}$) and by an increase of the width of the charge neutrality peak [33]. Finally, to identify the potential occurrence of a bandgap in graphene, we also performed temperature-dependent measurements (Figure 2e). Down to 7K no indications of a bandgap could be found, but only a decrease of the doping density.

Having established the 2DPI/graphene HS, we also have used the versatility of the chemical synthesis method to realize 2D polyimide films with varying thicknesses and transfer graphene onto them. In general, increasing the layer number of the 2DPI should increase the intralayer charge transfer due to an increased work function in the 2DPI as a function of layer thickness, as was seen in other 2D materials [34–37]. In fact, we do observe this in our transport measurements of graphene stamped on thicker 2DPI crystals (Figure S7), where in our accessible back gate window we are not able to electrostatically dope the graphene in the HS to the charge neutrality point, indicative of a doping density larger than ~ 5 x 10$^{12}$ cm$^{-2}$. These findings are consistent with our first-principle calculations. As shown in Figure 3a, adding more layers of the 2D polyimide leads to a substantial increase in charge transfer to graphene. It also leads to a significant down-shift of the Fermi level, which can be attributed to the p-doping of graphene (Figure S10). Our calculations show that the 2DPI has a higher work function than graphene; 4.56 eV versus 4.45 eV, which is consistent with the acceptor-character of the porphyrin sites [38–40], which also show a larger charge transfer as shown in Fig 1e and S11. These doping effects cannot be verified solely through transport measurements, as an accurate estimation of doping requires visualization of the Dirac peak. Therefore, an alternative method should be employed for comparing samples of different thicknesses, as described in the following.

**Spectroscopic analysis of doping and strain**

To compare the interaction effects across different types of samples and to obtain large sample statistics, we used Raman spectroscopy, as the scanning probe methods mentioned earlier only allow for relative comparisons within a single image. Furthermore, Raman spectroscopy allows for the distinction between doping and strain effects in the different HS samples. Strain is predicted to be present in the HS (Figure 1c) along with doping.

Typical Raman spectra of various 2DPI and 2DPI/graphene HSs are shown in Figures 3b-d (please note that in Figures 3b,c we have used 2DPI multilayers and in Figure 3d a 2DPI monolayer for the HS). One of the most striking effects is the enhancement of the peak intensities in the multilayer 2DPI/graphene HS compared to the bare 2DPI. In general, two main mechanisms are known that can lead to Raman enhancement in thin multilayers of this type, namely chemical enhancement and optical interference. The mechanism referred to as chemical enhancement includes various factors, such as charge transfer and orbital coupling [41–43]. Optical interference involves constructive interference of multiple reflections of the excitation light beam through the various layers of the sample [44, 45], and therefore is known as interference-enhanced Raman scattering, and depends critically on the dielectric constant of the layers involved. To unravel which mechanism underlies the Raman enhancement, we have investigated different types of samples: pure 2DPI, 2DPI which was protonated to mimic doping without the presence of additional layers, 2DPI/graphene and 2DPI/hBN, as shown in Figures 3 and S12. Additionally, we tested different thicknesses of the 2DPI film.

One method we have used to disentangle the effects stemming from the dielectric environment and charge transfer is to place a thin hBN flake (around 5 nm thick) and a graphene flake next to one another on the same thick 2DPI film. In both cases (in the 2DPI/graphene and 2DPI/hBN HSs) the Raman signal is enhanced by more than one order of magnitude, with distinct enhancement factors for different peaks. For instance, the enhancement factors for the peak at $\omega_1$=1301 cm$^{-1}$ are 21 and 32 for hBN and graphene HS respectively. One aspect contributing to the enhancement are interference effects. Specifically at the 532 nm excitation energy used in this experiment, the conditions to obtain an interference enhanced Raman scattered signal are met. In contrast, while with an excitation energy of 633 nm the 2DPI peaks are still enhanced in the 2DPI/graphene HS, they are not enhanced in the 2DPI/hBN HSs. We attribute this difference of enhancement to the dissimilar dielectric constant of hBN and graphene in the visible wavelength, with the consequence that no interference enhancement is taking place for the hBN HS (see Figure S12f). A second aspect is that the chemical enhancement, induced by the charge transfer between the 2DPI and the graphene, will also contribute to enhancing the Raman signal. A method to validate the impact of chemical enhancement to the Raman signal is to purposely dope (protonate) the pure 2DPI with hydrochloric acid. Indeed, protonation leads to an enhancement of the same Raman peaks (Figure S12c) as in the 2DPI/graphene HS. For instance, the peak at $\omega_2$=1381 cm$^{-1}$ is enhanced by a factor of 2. The normalized spectra presented in Figure 3c show that the relative intensity of different peaks undergoes similar changes in the doped 2DPI and also in the two HSs investigated.

We can further validate the relative roles of chemical and interference enhancement by investigating monolayer 2DPI HSs. The study of the HSs created with monolayer-2DPI (Figures 3d and S12b) indicates that chemical enhancement continues to play a role, while interference enhancement does not contribute significantly. We conclude this from a study performed with h-BN of different thicknesses on top of a 2DPI (Figure S12), where no enhancement can be observed. This lack of interference enhancement, combined with a smaller charge exchange in the monolayer case, results in significantly smaller enhancement in monolayer 2DPI/graphene HS compared to the thicker samples. Nevertheless, in the case 2DPI is protonated prior to HS formation, the peaks are even further enhanced compared to the undoped 2DPI/graphene HS (Figure 3d) since the charge charge transfer between the protonated 2DPI and the graphene increases in the doped HS.

This hierarchy of charge transfer (largest charge transfer in doped thick 2DPI / graphene HSs, smallest charge transfer undoped monolayer 2DPI/graphene HSs) can be further confirmed by the analysis of the relative intensity of graphene's main spectral features, the G ($\omega_G$~1580 cm$^{-1}$) and the 2D peak ($\omega_{2D}$~2680 cm$^{-1}$) [46, 47]. The value of I(2D)/I(G) is known to be about 4 for undoped graphene and to continuously decrease as the doping of the sample increases [48]. In our experiments, the relative intensity I(2D)/I(G) was calculated in multiples spots of differently prepared devices, as shown in Figure 3e. Here it is evident that the mean I(2D)/I(G) ratio of the bare graphene decreases in the undoped monolayer 2DPI/graphene HS, and it further decreases in the protonated monolayer-2DPI HS, reaching the lowest values when the HS is formed by thick crystals. The analysis of the spectral shape of the individual 2D and G peaks can allow to disentangle effects from doping and strain induced into graphene by the 2DPI. It is well-known that the Raman spectrum of graphene is highly sensitive to the mechanical strain and doping level of the flakes, as strain and doping both have a strong influence on the bond lengths and the electron-phonon coupling and directly impact the shape and the position of these two peaks. To test for the different contributions of strain and doping, the position of the 2D peak frequency $\omega_{2D}$ is plotted with respect to the G peak $\omega_G$ for multiple spots of differently prepared HSs in Figure 3f. By evaluating the slope of the experimental points with respect to the case of the unstrained and undoped graphene ($\omega_{G0}$ and $\omega_{2D0}$ indicated by the star), we disentangle the effects of the strain and doping [49, 50]. The 2 gray lines in Figure 3e separate the $\omega_G - \omega_{2D}$ plane in different regions. The gray area indicated with n=0, with a slope ($\Delta\omega_{2D}/\Delta\omega_G$) =2.2±0.2, represents the area for which the only effect induced on the graphene is strain. Starting from [$\omega_{G0},\omega_{2D0}$], the values move along the dotted line indicated with ε=0, characterized by a slope of ($\Delta\omega_{2D}/\Delta\omega_G$)=0.8, define a region in which the only effect induced on the graphene by the substrate is hole doping. By projecting the experimental values on the [$\omega_G,\omega_{2D}$] onto the 2 characteristic n=0 and ε=0 lines, it is possible to deduce the strain and doping of the studied sample: the further the projected point is from the unstrained and undoped value [$\omega_{G0}$, $\omega_{2D0}$] the higher the sample is doped/strained. By performing a linear fit on the experimental data, shown by faint gray lines on figure 3f, we obtained values of the slope of: $S_{Gr}$=-0.03± 0.26, $S_{Mono}$=1.24± 0.24, $S_{Thick\ cryst}$=0.94±0.28, $S_{Mono+}$=0.75±0.21. We therefore conclude that in all samples doping plays the main role since the 2D peaks of the HSs are located mostly parallel to the ε=0 line. The doping density (as deduced from the G-peak position) increases according to the doping hierarchy discussed above. Finally, while some strain seems to present in the monolayer 2DPI / graphene samples, it overall plays a minor role compared to the doping.

**Direct analysis of graphene band structure in 2DPI/graphene HS by ARPES**

Up to now the analysis has indicated that the 2DPI induces interlayer charge transfer, charge scattering and strain in the HS, but we have not been able to experimentally identify the expected change in the density of states in the HS. To test for these effects, we have manufactured HSs directly on conductive substrates (doped Si) to allow direct band structure measurements by angle-resolved photoemission spectroscopy (ARPES) (technique described in [51, 52]). In the momentum cut of the detected spectra, shown in Figure S13b, it is possible to identify that there is a dip in the acquired spectral weight at the Dirac peak. This spectral weight dip of about 215 meV is consistent with the dip in the density of states in the HS as calculated in the substrate-supported HS (Figure 1d, where two distinct density peaks emerge below and above the Fermi level). The dip in the ARPES spectral weight is therefore a sign of the strong interaction between graphene and 2DPI, more details of the analysis are shown in Figure S13 in the SI. There, we also discuss that interaction between 2DPI and graphene when placed on a Si substrate is stronger than when $SiO_2$ substrates are used, as confirmed by our Raman analysis and theoretical calculations (Figure S6b).

**Towards larger device sizes: 2DPI/ CVD graphene heterostructures**

In the measurements up to now exfoliated graphene has been used, which is ideal for fundamental studies. However, in such devices the HS size is limited by the size of graphene to the μm regime. Since the 2DPI monolayers can be also synthesized on larger scales, we have tested HS composed of chemical vapor deposition (CVD) graphene and the 2DPI (Figure S14). In these samples similar inter-layer charge transfer between the 2D polyimide and the graphene was measured, whereas the overall charge carrier mobility was lower, as expected for CVD graphene.

**Conclusion**

In conclusion, we conducted an in-depth study of the interaction between a two-dimensional covalent organic framework and graphene, using various experimental techniques and density functional theory calculations. Through electronic, Kelvin probe force microscopy and Raman spectroscopic measurements we showed that graphene is hole doped by the 2D polyimide. This charge transfer process is likely attributed to the work function difference between the 2DPI and graphene. Our investigation of different samples revealed that controlling the charge transfer between the polymer and graphene can be finetuned by adjusting the thickness of the 2DPI and/or protonating the 2DPI. Specifically, as the 2D polyimide thickens, the hole doping effect on graphene increases. The remarkable tunability observed in the interaction between this 2D polyimide and graphene suggests a promising avenue for further exploration. With diverse 2D polymers characterized by distinct chemical and topological properties, we anticipate the investigation of new and intriguing physical phenomena in this emerging field of study.

**Experimental methods**

**Device Fabrication:**

The stamped samples were fabricated by transferring the graphene flakes on the 2DPI substrate using the dry transfer method described in [19]. The graphene flakes were obtained through mechanical exfoliation from natural graphite crystals (from NGS trading and consulting) Silicon/Silicon dioxide (300 nm) substrate. The electrical contacts were patterned using electron beam lithography (from Raith), with the following parameters: an accelerating voltage of 10kV, a dose of 110 $\mu$C cm$^{-2}$ for the 7.5 $\mu$m aperture (used for small contacts) and a dose 170 $\mu$C cm$^{-2}$ for 60 $\mu$m (used for wider contact lines). The layer of resist for the e-beam procedure was obtained following the procedure described in [53]. Finally, the 1 nm chromium (with a rate of around 0.43 Å/s) and 60 nm gold (with a rate of around 0.9 Å/s) contacts are evaporated via thermal evaporation (evaporation chamber from BesTec) at pressures of around 10$^{-6}$ mbar. The top electrolyte gate for the measurements, shown in figure 2, was deposited through the technique described in ref [54]. To separate the CVD graphene and the thick crystals samples from the surroundings, the flakes were etched through a dry etching process performed with a flow of 40 sccm O$_2$ plasma at 80 W and 40 mTorr for 18 seconds in a reactive ion etching chamber (from Oxford PlasmaLab). The etching masks were designed using electron beam lithography. In the protonated sample the protonation was performed by depositing a droplet of 10% of Hydrochloridric acid and letting it dry at 80° overnight (around 12 h).

Before preparing the samples, the stability of the studied 2DPI to the applied chemicals was assessed through the analysis of AFM pictures and Raman spectra; no difference was found before and after the application of any of the used chemicals.

**Scanning probe techniques:**

The AFM and KPFM measurements were performed with Asylum Jupiter AFM by Oxford Instruments. The tips used are the Tap300Al-G (from NanonAndMore) for AFM topography and SCM-PIT (from NanonAndMore) for the KPFM measurements. The KPFM mode used was a two-pass lift mode in which the sample is scanned twice. In the first scan, the topography of the sample is detected, while in the second scan only the surface potential is detected. Our measurement technique is an AM-KPFM, meaning that the surface potential $U_{CPD}$ is measured by controlling the oscillating amplitude of the cantilever. In these types of measurements, an AC bias, characterized by a frequency ω, is applied to the conductive tip, producing a force F between the tip and the sample. Modelling the force like parallel plate capacitor, we obtain:

$$F = \frac{1}{2}\frac{\partial C}{\partial z}V^2$$

Where C is the capacitance between the tip and the sample, z is the vertical distance between the two and V is the total potential difference between the tip and the probe. The potential V is given by the applied AC potential $V_{AC}$, the contact potential $U_{CPD}$ we would like to measure and an additional DC voltage $V_{DC}$ that needs to be externally applied during measurements:

$$F = \frac{1}{2}\frac{\partial C}{\partial z}\left(\left[(V_{DC} - U_{CDP})^2 + \frac{1}{2}V_{AC}^2\right] + 2[(V_{DC} - U_{CDP})V_{AC}sin(\omega t)] - \left[\frac{1}{2}V_{AC}^2 cos(2\omega t)\right]\right)$$

By locking the amplitude to the frequency ω with a lock-in amplifier, the minimization of the amplitude is obtained applying an additional DC equal to the surface potential of the probe, $V_{DC} = U_{CDP}$. The $U_{CDP}$ is therefore measured with a feedback loop calibrated to adjust the DC bias applied to the lever in order to minimize the amplitude.

The near field scattering microscope images are taken with a commercial s-SNOM (from Neaspec Company) coupled to a tunable $CO_2$ laser (from Access laser, model L4G) with wavelengths of 9.2–10.78 µm. The infrared nanoimaging was based on an atomic force microscopy (AFM) operated in tapping mode with a tapping amplitude of Δ$z$=90 nm. All the images were taken with the tips Arrow-NCPt (from Nanoworld) characterized by a tapping frequency Ω of ~270 KHz. The power of the laser during the measurement was set to around 0.8 mW.

**Raman measurements:**

The Raman measurements shown in the manuscript were taken with the commercially available setup is the commercially available LabRam HR Evolution (from Horiba). This setup is coupled two 2 lasers: an HeNe laser with a 632,9 nm and YAG-Laser (Neodymium-doped Yttrium Aluminum Garnet) with a wavelength of 532 nm (torus 532 from Laser Quantum). The setup is equipped with 2 different gratings of 1800 gr/mm and 600 gr/mm, both were used during this work. The images shown in Figure 4 are taken with the 600 gr/mm grating.

**Electrical measurements:**

Electrical measurements, both at room temperature and low temperature, were performed by contacting the gate and the source-drain of the samples through the application of voltages to needles connected to a Keithley 2450. In measurements involving the electrolyte gate, the needles were applied directly to the droplet of electrolyte on top of the sample. The measurements under vacuum were conducted in a Lakeshore CRX-VF probe station under vacuum conditions (temperature range 5 K–450 K). The samples were fixed to the sample holder using silver conducting paint to ensure thermal connection between the holder and the sample.

From gate sweep measurements shown in Figure 2c, considering the field effect transistor geometry of the samples it is possible to derive the charge carrier density $n$ induced capacitively by the Si gate, with:

$$n = \frac{\varepsilon_0 \varepsilon}{ed}(V_{BG} - V_0)$$

where d is the thickness of the SiO$_2$ (300 nm), is the dielectric constant of the dioxide, is the vacuum permittivity, $e$ is the elementary electron charge, $V_{BG}$ is the applied voltage to the back-gate and $V_0$ is the voltage corresponding to the CNP of the considered sample [55]. The mobility, corresponding to a defined $n$, was calculated from the 2 point-probe resistivity $\rho$ from: $\mu=1/(\rho n e)$ .

**Quantum chemistry calculations details:**

All structures are generated by hetbuilder [56] by rotating and expanding unit cells of target 2D-polymer and graphene w/wo substrate to look for the shared coincident supercell. The geometries of all multi-layer structures were optimized by Density Functional based Tight Binding method [57, 58] (DFTB), which is a computationally efficient Tight Binding approach based on density functional theory. DFTB+ [59], as an implementation of DFTB, was used to perform the geometry optimizations by matsci-0-3 parameters [60]. Electronic properties like band structures and density of states were performed by Fritz-Haber-Institute ab-intitio materials simulations package (FHI-aims [61]) with PBE [62] functional plus Many-Body dispersion (MBD) [63]. Tier 2 basis set and tight integration mesh were used. Charge transfer calculations were performed using Vienna Ab initio Simulation Package (VASP) [64–66] with PBE functional plus D3BJ (D3 with Becke-Johnson damping) dispersion [67] and Bader charge analysis was done by code from Dr. Henkelman's group [68–70]. Strain analysis was done by a self-made script using Atomic Simulation Environment (ASE) [71], which can be found in Github (https://github.com/shuangjiezhao).

**Acknowledgement:**

We acknowledge stimulating discussions with Stephanie Reich and Sabrina Jürgensen. F.F., J.P., R. T. W., X.F. and TH acknowledge funding from the SPP 2244 (2DMP) and from the SPP 1928 (COORNETs). S. Z. acknowledges funding of the DFG priority program SPP 2244. The authors gratefully acknowledge the computing time made available to them on the high-performance computers at the NHR Centers at TU Dresden and NHR Center PC2. These are funded by the Federal Ministry of Education and Research and the state governments participating on the basis of the resolutions of the GWK for the national high-performance computing at universities (www.nhr-verein.de/unsere-partner). K.W. and T.T. acknowledge support from the JSPS KAKENHI (Grant Numbers 20H00354, 21H05233 and 23H02052) and World Premier International Research Center Initiative (WPI), MEXT, Japan.

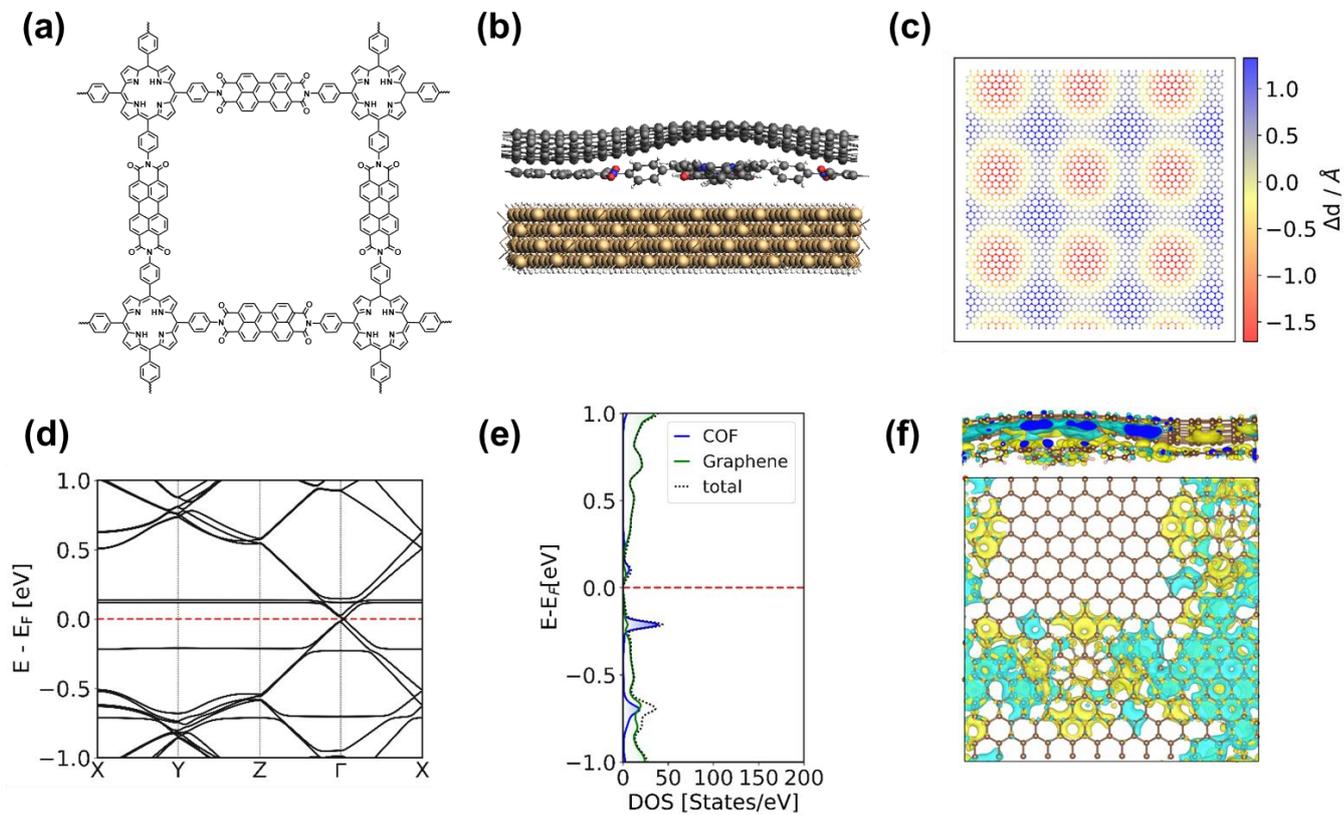

*Figure 1 (a) Chemical structure of studied 2DPI. (b) Atomic structure of Si/2DPI/graphene. (c) the corrugation of graphene induced by 2D-polymer and substrate Si/2DPI/graphene (Red: graphene bent towards polymer, Blue: graphene bent from polymer), forming square superlattice. (d1) band structure of 2DPI/graphene moiety of Si/2DPI/graphene structure. (d2) Projected Density of States of 2DPI/graphene moiety of Si/2DPI/graphene structure. (e) Visualization of charge distribution difference 2DPI/graphene moiety Si/2DPI/graphene. (Yellow: charge accumulation. Light blue: charge depletion)*

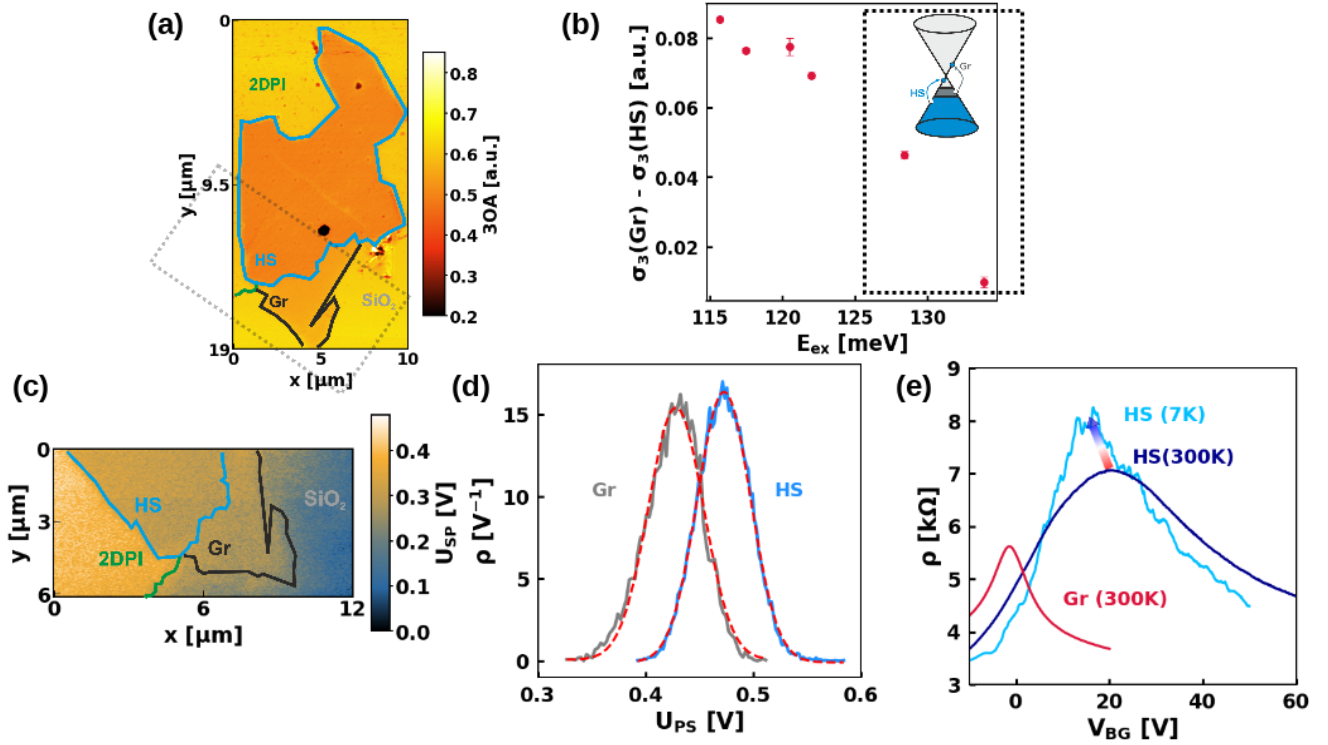

*Figure 2 Interlayer charge transfer in monolayer-monolayer 2DPI/graphene HS. (a) 3rd harmonic near field optical amplitude at 116 meV of one of the samples studied. In this sample the graphene was transferred through dry transfer method on the 2DPI. (b) Energy dependent difference between the normalized 3rd harmonic optical signal of the graphene and the HS part of the samples shown in S4a, where the optical signal was normalized to the value of the gold marker. In the insets, the process associated to the higher excitation energies is shown: both the graphene and the HS are dominated by inter-band process. (c) KPFM measurement of a sample exfoliated on a monolayer of 2D polyimide. (d) The gaussian distribution density of the $U_{SP}$ values obtained for the HS and the bare graphene part of the image. From the $U_{SP}$ difference between the bare graphene part and the HS part we could estimate the hole doping induced by the 2DPI. The dotted red line shows the gaussian fits performed on the distributions. (e) Gate sweep measurement over the $Si^+$ back gate sample prepared on a $SiO_2/Si^+$ wafer. The measurement is performed under vacuum. The HS show hole doping in respect of the bare graphene part of the sample.*

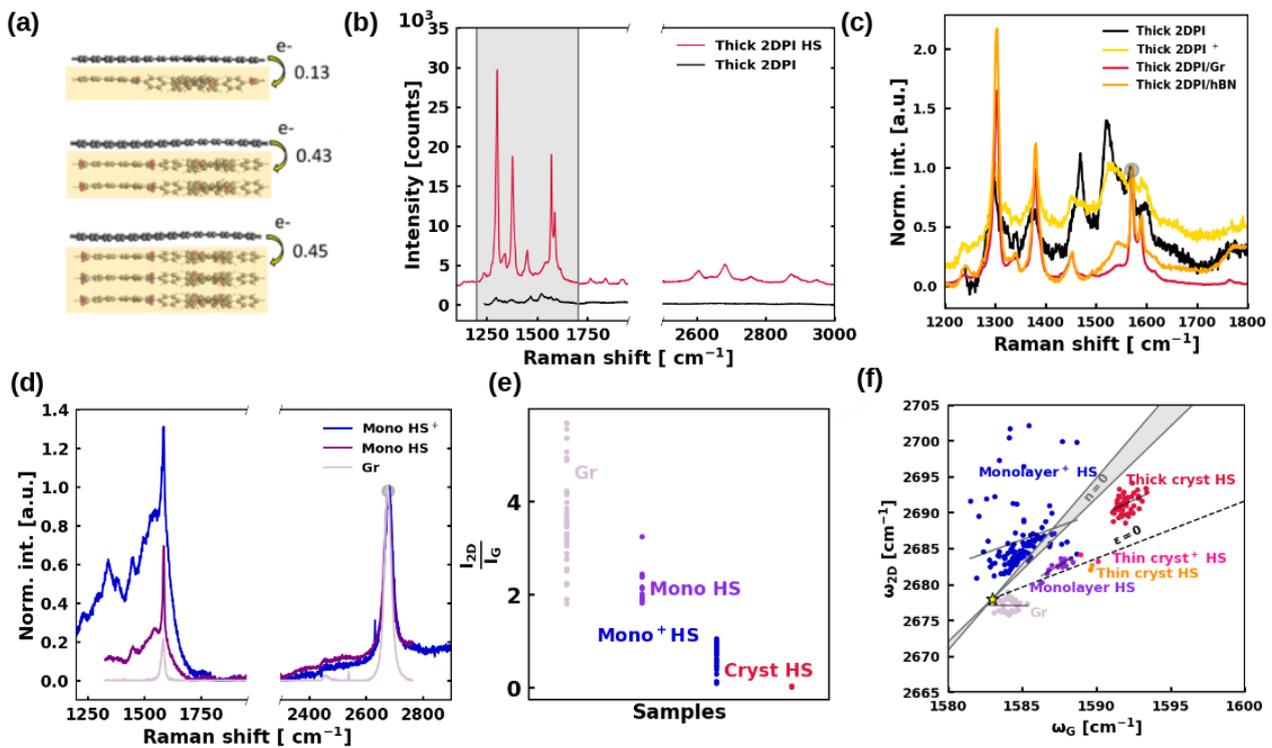

*Figure 3: (a) Representation of graphene HS with one, two or three layers of 2D polyimide and net electron transfer per unit cell (b) Comparison of the Raman spectra taken on a sample with thick crystals: in black the spectrum of the untreated crystals, and in red the one of the thick 2DPI/graphene HS. To both spectrum the background has been subtracted and an to the graphene one offset of a multiple of 2500 has been added in order to better visualize each individual one. The offset has been added in order to maintain the hierarchy of the most counts measured. The range indicated in the shaded gray region is shown in figure (c) where the shown spectra where normalized in respect of the peak at ω~1570 cm$^{-1}$, circled in the figure, which is very prominent in each spectrum. Here there are 2 additional spectra in respect of figure 2a: in yellow is shown the one of the crystals protonated with a in orange the spectrum of the 2DPI/hBN stack sample (d) Comparison of the Raman spectra taken 2 samples in which the graphene is placed on top of a 2D polyimide monolayer. The pink and purple spectra are taken from the same sample respectively on the HS and the bare graphene part. The spectrum in blue belongs to a sample in which the 2DPI has been protonated before the graphene transfer. (e) Ratio between the intensity of the 2D peak and the G peak of different graphene samples. the fact that the ratio decreases both with the protonation and the presence of the thicker 2DPI, confirms the hypothesis that the charge transfer between the two materials increases with the increasing number of polyimide layers underneath and with protonation of COF upon stamping. (f) Raman spectra analysis comparing different sample. Here the $ω_G$ is plotted in respect of $ω_{2D}$ all the different types of sample studied. The spots mainly lay parallel to the line corresponding to null strain (ε=0), showing that the doping is stronger effect than the strain on the graphene. The shift of the $ω_G$ to higher values with thicker samples, support the hypothesis that the doping increases with increases 2DPI thickness.*

17